\documentclass[floatfix,aps,prl,showpacs,amsmath,nofootinbib,
preprintnumbers,twocolumn]{revtex4}
\usepackage{graphicx}

\begin{document}

\preprint{FERMILAB-PUB-05-024-A}

\title{Primordial inflation explains why the universe is accelerating today}
\author{Edward W. Kolb}\email{rocky@fnal.gov}
\affiliation{Particle Astrophysics Center, Fermi
       	National Accelerator Laboratory, Batavia, Illinois \ 60510-0500, USA \\
       	and Department of Astronomy and Astrophysics, Enrico Fermi Institute,
       	University of Chicago, Chicago, Illinois \ 60637-1433 USA}

\author{Sabino Matarrese}\email{sabino.matarrese@pd.infn.it}
\affiliation{Dipartimento di Fisica ``G.\ Galilei,'' Universit\`{a} di Padova, 
        and INFN, Sezione di Padova, via Marzolo 8, Padova I-35131, Italy}

\author{Alessio Notari}\email{notari@hep.physics.mcgill.ca}
\affiliation{Physics Department, McGill University, 3600 University Road,
        Montr\'eal, QC, H3A 2T8, Canada}

\author{Antonio Riotto}\email{antonio.riotto@pd.infn.it}
\affiliation{INFN, Sezione di Padova, via Marzolo 8, I-35131, Italy}

\date{\today}

\begin{abstract}
We propose an explanation for the present accelerated expansion of the universe
that does not invoke dark energy or a modification of gravity and is firmly
rooted in inflationary cosmology.
\end{abstract}


\maketitle


In recent years the exploration of the universe at redshifts of order unity has
provided information about the time evolution of the expansion rate of the
universe.  Observations indicate that the universe is presently undergoing a
phase of accelerated expansion \cite{acceleratedreview}. The accelerated
expansion is usually interpreted as evidence either for a ``dark energy'' (DE)
component to the mass-energy density of the universe or a modification of
gravity at large distance.  The goal of this {\it Letter} is to provide an
alternative explanation for the ongoing phase of accelerated expansion that is,
we believe, rather conservative and firmly rooted in inflationary cosmology. 

In the homogeneous, isotropic Friedmann-Robertson-Walker (FRW) cosmology, the
deceleration parameter $q$ describes the deceleration of the cosmic scale
factor $a$.  It is uniquely determined by the relative densities and the
equations of state of the various fluids by (overdot denotes a time derivative)
\begin{equation}
\label{qdef}
q \equiv -\frac{\ddot{a}a}{\dot{a}^2}
=\frac{1}{2}\Omega_\textrm{TOT}+\frac{3}{2}\sum_i\,w_i\,\Omega_i ,
\end{equation}
where $\Omega_\textrm{TOT}$ is the total energy density parameter and the
factors $\Omega_i$ are the relative contributions of the various components of
the energy density with equation of state $w_i=p_i/\rho_i$ ($p_i$ and $\rho_i$
are the pressure and energy density of fluid $i$). The expansion accelerates if
$q<0$. Observations seem to require DE with present values $w_{DE}\sim -1$ and 
$\Omega_{DE}\sim 0.7$ \cite{rp}.  The negative value of $w_{DE}$ is usually
interpreted as the effect of a mysterious fluid of unknown nature with negative
pressure or a cosmological constant of surprisingly small magnitude.

Our proposal is as follows. Suppose cosmological perturbations with wavelengths
larger than the present Hubble radius, $H_0^{-1}$, exist. A local observer
inside our Hubble volume would not be able to observe such super-Hubble modes
as real perturbations. Rather, their effect would be in the form of a classical
(zero-momentum) background. Suppose further that our local universe is filled
with nonrelativistic matter and no DE. We show that if the long-wavelength
perturbations evolve with time, a local observer would infer that our universe
is not expanding as a homogeneous and isotropic FRW matter-dominated universe
with Hubble rate $H=\frac{2}{3}t^{-1}$, where $t$ is cosmic time. On the
contrary, the universe would appear to have an expansion history that depends
on the time evolution of the super-Hubble perturbations. Potentially, this
could lead to an accelerated expansion. 

The origin of the long-wavelength cosmological perturbations is inflation. 
Inflation is an elegant explanation for the flatness, horizon, and monopole
problems of the standard big-bang cosmology \cite{lrreview}.  But perhaps the
most compelling feature of inflation is a theory for the origin of primordial
density perturbations and anisotropies in the cosmic microwave background
(CMB).  Density (and gravitational-wave) perturbations are created during
inflation from quantum fluctuations and redshifted to sizes larger than the
Hubble radius. They are then ``frozen'' until after inflation when they
re-enter the Hubble radius.

A consequence of inflation is scalar perturbations of wavelength larger than
the Hubble radius.  During inflation a small region of size less than the
Hubble radius grew to encompass easily the comoving volume of the entire
presently observable universe. This requires a minimum number of
\textit{e}-foldings, $N\gtrsim 60$, where $N$ measures the logarithmic growth
of the scale factor during inflation. Most models of inflation predict a number
of \textit{e}-foldings that is, by far, much larger than 60 \cite{lrreview}.
This amounts to saying that today there is a huge phase space for super-Hubble
perturbations.  These super-Hubble perturbations will re-enter the Hubble
radius only in the very far future. But if they evolve with time, we will
demonstrate that they will alter the time evolution of the Hubble rate
experienced by a local observer.

Let us consider a universe filled only with nonrelativistic matter. Our
departure from the standard treatment is that we do not treat our universe as
an idealized homogeneous and isotropic FRW model, rather we account for the
presence of cosmological perturbations. We work in the synchronous and comoving
gauge and write the line element as 
\begin{equation}
ds^2 = -dt^2 + a^2(t) e^{-2\Psi(\vec{x},t)}\delta_{ij}dx^idx^j , 
\label{oo} 
\end{equation}
where $t$ is the cosmic proper time measured by a comoving observer and
$\Psi(\vec{x},t)$ is the gravitational potential. We have neglected the
traceless part of the metric perturbations, which contains a scalar degree of
freedom as well as vector and tensor modes, because they will not play a
significant role in what follows (even if they must be taken into account
consistently when solving Einstein's equations). Perturbations $\Psi$
correspond to a local conformal stretching or contracting of Euclidean
three-space.

How should one deal with perturbations that have wavelengths larger than the
Hubble radius?  As we have already stressed, a local observer would see them in
the form of a classical zero-momentum background. This suggests that
super-Hubble modes should be encoded in the classical (zero-momentum) scale
factor by a redefinition of the metric. To do so, we split the gravitational
potential $\Psi$ into two parts: $\Psi = \Psi_\ell + \Psi_s$, where
$\Psi_\ell$ receives contributions only from the long-wavelength super-Hubble
modes and $\Psi_s$ receives contributions only from the short-wavelength
sub-Hubble modes. 

By construction, $\Psi_\ell$ is a collection of Fourier modes of wavelengths
larger than the Hubble radius, and therefore we may safely neglect their
spatial gradients within our Hubble volume.  This amounts to saying that
$\Psi_\ell$ is only a function of time. We may then recast the metric of Eq.\
(\ref{oo}) within our Hubble volume in the form
\begin{equation}
ds^2 = -dt^2 + \overline{a}^2(t)\, e^{-2\Psi_s}\delta_{ij}dx^idx^j , 
\label{ooo}
\end{equation} 
where we have defined a new scale factor $\overline{a}$ as
\begin{equation}
\label{defabar}
\overline{a}(t) = a(t) \, e^{-\Psi_\ell(t) + \Psi_{\ell 0}}.
\end{equation}
(Here and below, a subscript ``$0$'' denotes the present time.)  We have taken
advantage of the freedom to rescale the scale factors to set
$\overline{a}_0=a_0=1$.  Our original perturbed universe is therefore
equivalent to a homogeneous and isotropic universe with scale factor
$\overline{a}$, plus sub-Hubble perturbations parametrized by $\Psi_s$, which
we shall ignore. A local observer, restricted to observe within our Hubble
volume, would determine an expansion rate 
\begin{equation}
\label{h}
\overline{H} \equiv \frac{1}{\overline{a}}\frac{d\overline{a}}{dt} 
= H - \dot{\Psi}_\ell .
\end{equation} 

Notice that Eq.\ (\ref{h}) does not coincide with the expansion rate of a
homogeneous and isotropic matter-dominated FRW universe \textit{if the
cosmological perturbations on super-Hubble scales are time dependent.}  Indeed,
if $\Psi_\ell$ is constant in time, it may be eliminated from the metric of
Eq.\ (\ref{oo}) by a simple rescaling of the spatial coordinates while still
remaining in the synchronous gauge. But the important point is that the freedom
to rescale the spatial coordinates while remaining in the synchronous gauge is
lost if the inhomogeneities have a non-trivial time dependence on super-Hubble
scales.  Similarly, a local observer restricted to live within a single Hubble
volume will measure a deceleration parameter
\begin{equation}
\overline{q}=-1+\frac{3/2+\ddot{\Psi}_{\ell}/H^2}{\left(1-
\dot{\Psi}_\ell/H\right)^2} ,
\end{equation}
which---because of the action of time-dependent super-Hubble modes---deviates
from the value predicted in a homogeneous and isotropic flat matter-dominated
universe ($\overline{q}=1/2$). 

What then is the expected value of the deceleration parameter?  Solving
Einstein's equations, we find that (up to higher derivative terms) at any order
in perturbation theory the metric perturbation assumes the form 
\begin{equation}
\label{np}
\Psi(\vec{x},t) = \frac{5}{3}\varphi(\vec{x}) + a(t) \frac{2}{9} 
\frac{e^{10\,\varphi(\vec{x})/3}\nabla^2 \varphi(\vec{x})}{H_0^2}  ,
\end{equation}
where we also used the fact that $(\nabla \varphi)^2 \ll |\nabla^2 \varphi|$.
Eq.\ (\ref{np}) reproduces the evolution of the cosmological perturbations
generated during a primordial epoch of inflation in a matter-dominated universe
at second order found in Ref.\ \cite{KMNR}.  Here, $\varphi(\vec{x})$ is the
so-called peculiar gravitational potential related to the density
perturbations, and the values of the potentials have been computed with a
proper match to the initial conditions set by single-field models of inflation
\cite{en}. Of particular interest to us is the infrared part of $\Psi_\ell$. 
As shown in Ref.\ \cite{KMNR}, under mild assumptions about the spectrum of
super-Hubble perturbations, terms like ${\rm
exp}(10\,\varphi/3)\nabla^2\varphi$ can be large (order unity) due to the large
variance in $\varphi$ already at second-order in perturbation theory.  As
discussed, the time-independent part of $\Psi_\ell$ can be absorbed into a
rescaling of coordinates.  We may therefore write
\begin{equation}
\Psi_\ell = a(t) \left[ \frac{2}{9} 
\frac{e^{10\,\varphi/3}\nabla^2 \varphi}{H_0^2} \right] 
\equiv a(t) \ \Psi_{\ell 0} .
\end{equation}

From Eqs.\ (\ref{h}-\ref{np}), one may easily compute the super-Hubble
contributions to the expansion rate and the deceleration parameter.  
Using the fact that $a\propto t^{2/3}$, we obtain
\begin{eqnarray}
\label{hofa}
\overline{H} & = & \frac{\overline{H}_0}{1-\Psi_{\ell 0}}
\left( a^{-3/2} - a^{-1/2} \ \Psi_{\ell 0} \right) , \\
\overline{q} & = & -1 +\frac{3/2 - a \ \Psi_{\ell 0}/2}
{\left(1 - a \ \Psi_{\ell 0} \right)^2} ,
\label{qbar}
\end{eqnarray}
which at second order nicely reproduces the findings of Refs.\ \cite{KMNR} and
\cite{bmr} for the contribution from the super-Hubble modes to the expansion
rate and deceleration parameter, respectively.  In a universe filled only with
nonrelativistic matter we see that the deceleration parameter is not uniquely
determined to be 1/2!  One must account for the statistical nature of the
vacuum fluctuations from which the gravitational potential originated. The
gravitational potential does not have well defined values; one can only define
the probability of finding a given value at a given point in space. Our Hubble
volume is just one of many possible statistical realizations, and the present
deceleration parameter is expected to deviate from 1/2 by an amount roughly
$\langle \Psi_{\ell 0} \rangle \sim 
\sqrt{\textrm{Var}\left[\exp(10\,\varphi/3)\nabla^2\varphi \right]}/H_0^2$
(here $\langle\cdots\rangle$ denotes the spatial average in our Hubble volume).
The variance must be computed using the statistical properties of the
perturbations, namely the primordial power spectrum. The computation of the
variance involves a potentially large infrared contribution that depends upon
the total number of \textit{e}-foldings and the spectral index $n$ of the
primordial power spectrum. For $n\leq1$, a variance of order unity may be
obtained \cite{KMNR}. Also note that the departure from $\overline{q}=1/2$
increases with time, so if the departure is of order unity today, it was
irrelevant early in the expansion history of the universe. On the other hand,
if we extend the validity of Eq.\ (\ref{qbar}) into the far future, the
deviation of the deceleration parameter from $\overline{q}=1/2$ becomes larger
and larger, approaching the asymptotic value $\overline{q}=-1$.

The statistical nature of the effect implies that it is not possible to predict
the value of $\overline{q}$; however, we can say something about its variance
\cite{KMNR}.  Following Ref.\ \cite{KMNR}, for the infrared part we have
$\textrm{Var}[\exp(10\,\varphi/3)\nabla^2\varphi] \simeq
\textrm{Var}[\exp(10\,\varphi/3)]\textrm{Var}[\nabla^2\varphi]$, and within our
Hubble volume we find
\begin{equation}
\frac{\textrm{Var}[\nabla^2\varphi]\textrm{Var}\left[e^{10\varphi/3}
\right]} {H_0^4} \simeq I_3\,e^{50I_{-1}/9} ,
\end{equation}
where $I_i \equiv A^2 \int_{x_\textrm{MIN}}^1 dx \ x^{i-\epsilon} \ e^{-x^2}$.
The factor $A\simeq 2\times 10^{-5}$ is the perturbation amplitude on the scale
of the Hubble radius. We have assumed a scalar perturbation spectral index
$n=1+\epsilon$ for super-Hubble perturbations and $x_\textrm{MIN}$ is an
infrared cutoff which depends on the total number of $e$-folds of primordial
inflation (see Ref.\ \cite{KMNR} for a discussion). The factor $I_3$ is of
order $A^2$ and insensitive to $\epsilon$ and $x_\textrm{MIN}$, but if
$\epsilon \leq 0$, $I_{-1}$ is potentially large and sensitive to $\epsilon$
and $x_\textrm{MIN}$.  As discussed in Ref.\ \cite{KMNR}, it is easy to imagine
that $\sqrt{\textrm{Var}[\exp(10\,\varphi/3)]}$ is large enough for
$\exp(10\,\varphi/3)\nabla^2\varphi/H_0^2$ to be of order unity in our Hubble
volume.

One might worry that a large value of $\textrm{Var}[\exp(10\,\varphi/3)]$ might
indicate a breakdown of perturbation theory. Happily, that is not the case
since the effect depends on the cross-talk between small-wavelength
perturbations contributing to $\nabla^2\varphi$ and long-wavelength
perturbations contributing to $\varphi$.  The constant contributions to
$\Psi_\ell$ from $\varphi$ has already been absorbed into $\overline{a}$. 
Therefore, $\varphi$ only enters multiplied by the small value of
$\nabla^2\varphi$, i.e., the perturbation variable is really
$\exp(10\,\varphi/3) \nabla^2\varphi$.  So as long as $\left|\Psi_\ell\right|$
is less than unity, perturbation theory should be valid.  We will show that
values $\left|\Psi_{\ell0}\right| \simeq 0.5$ seem to be consistent with
observations, so perturbation theory is reasonably well under control.

The statistical nature of the perturbations also implies that a local observer
restricted to our local Hubble volume might observe a negative value of the
deceleration parameter even if the universe only contains nonrelativistic
matter. The underlying physical reason for this is that a primordial epoch of
inflation generated cosmological perturbations of wavelengths much larger than
the Hubble radius, and the perturbations evolve coherently with time and
influence the time evolution of the local Hubble rate. In other words, if
inflation took place, the correct solution of the homogeneous mode in a FRW
cosmology with nonrelativistic matter is not provided by the scale factor
$a(t)\sim t^{2/3}$, but by the scale factor $\overline{a}$. A na\"{\i}ve
cosmologist knowing nothing about the energy content of the universe and the
presence of the super-Hubble modes might (incorrectly) ascribe the accelerated
expansion to a fluid with negative pressure.  Comparing Eqs.\ (\ref{qdef}) and
(\ref{qbar}), super-Hubble perturbations mimic a fictitious DE fluid with
equation of state
\begin{equation}
\label{womegade}
w_{DE} \Omega_{DE} = \frac{2}{3}\left(\overline{q}-\frac{1}{2}\right).
\end{equation}

Extrapolating into the far future, Eq.\ (\ref{womegade}) tells us that
super-Hubble perturbations will mimic a pure cosmological constant!  Of course,
such an extrapolation is not entirely justified since the explicit expression
of Eq.\ (\ref{np}) is only valid up to higher-order derivatives. However, our
conclusions are supported by generic and well known results in general
relativity.  An underdense region of the universe (corresponding to our
accelerating Hubble volume), expands faster and faster, wiping out the initial
anisotropies (shear), eventually reaching an asymptotic stage of free
expansion, thus decoupling from the surrounding regions which continue to
expand as in a matter-dominated universe \cite{ibbmp}. Therefore, if the effect
of the super-Hubble modes is to render our local Hubble region underdense, a
local observer will perceive a faster than mean expansion as a net acceleration
whose asymptotic value of the deceleration parameter is precisely $-1$. 

In conclusion, we have proposed an explanation for the presently observed
acceleration of the expansion of the universe that is, in many ways, very
conservative.  It does not involve a negative-pressure fluid or a modification
of general relativity.  The basic point is that inflation naturally leads to a
large phase space for super-Hubble perturbations.  We claim that these
super-Hubble perturbations should properly be encoded into a new scale factor. 
The time evolution of the super-Hubble perturbations (i.e., the growth of
density perturbations) modifies the expansion rate of the new scale factor from
that of a flat matter-dominated FRW model. We propose that this modification of
the expansion rate is what is usually (improperly) ascribed to dark energy.  In
the far future it will lead to an expansion seeming to arise from a pure
cosmological constant, although the vacuum energy density will remain zero. 
Since inflationary super-Hubble perturbations are the key ingredient, we refer
to our model as ``Super-Hubble Cold Dark Matter'' (SHCDM).

There is only one parameter in SHCDM, $\Psi_{\ell 0}$.  It should be
regarded as a Gaussian random variable.  Its mean is zero, but within any
Hubble volume its variance (which can be large) may be expressed as an integral
over the super-Hubble perturbation power spectrum.  In principle, it is
completely specified by the physics of primordial inflation. If the
super-Hubble spectrum of perturbations is no bluer than the Harrison-Zel'dovich
spectrum, a reasonable number of $e$-folds of inflation will result in a
variance that is large enough so that acceleration would result.  Furthermore,
since super-Hubble perturbations grow with time, the observational evidence
that the universe is entering an accelerated phase just now may be ascribed to
the fact that super-Hubble perturbations have needed a sufficient amount of
time to grow starting from the initial conditions set by inflation.

\begin{figure}
\includegraphics[width=0.48\textwidth]{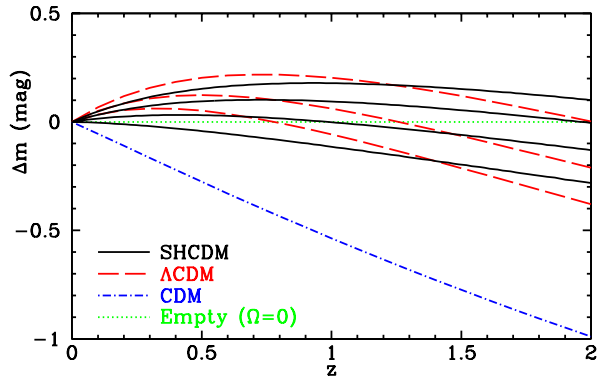}
\caption{\label{dm} The apparent magnitude difference as a function of redshift
between SHCDM and $\Lambda$CDM models compared to a model empty universe.  The
SHCDM models from top to bottom are for $\Psi_{\ell 0}= -1.0,\ -0.75,\ -0.5,\
\textrm{and}\ -0.25$, while the $\Lambda$CDM models from top to bottom are for
$\Omega_\Lambda=$ 0.8, 0.7, and 0.6 (all with $w=-1$). Also indicated is the
CDM model ($\Omega_\Lambda=0$).}
\end{figure}

Of course, observation is the ultimate arbiter between explanations for the
acceleration of the universe.  One observational test is the
luminosity-distance--redshift relationship.  The luminosity distance is given
by $d_L=(1+\overline{z})\int_0^{\overline{z}}dz'/\overline{H}(z')$.  In SHCDM
$\overline{H}$ is given by Eq.\ (\ref{hofa}), and the relationship between $a$
and $\overline{z}$ can be found from Eq.\ (\ref{defabar}).  It is traditional
to express the luminosity distance in terms of the apparent magnitude, $m=5\log
d_L +c$, where $c$ is an irrelevant constant.  Fig.\ \ref{dm} indicates that
SHCDM with $-0.25 \gtrsim \Psi_{\ell 0} \gtrsim -1$ is at present
indistinguishable from $\Lambda$CDM models. Future precision determinations of
the luminosity-distance as a function of redshift, as well as other precision
cosmological observations, should be able to discriminate between SHCDM and
other explanations for the acceleration of the universe.

It is widely believed that the acceleration of the universe heralds the
existence of new physics.  If our proposal is correct this is not the case.  On
the other hand, observational information about the acceleration of the
universe sheds light on the nature of primordial inflation and provides
precious knowledge of the workings of the universe beyond our present horizon.

E.W.K.\ is supported in part by NASA grant NAG5-10842 and by the Department of
Energy. S.M.\ acknowledges INAF for partial financial support



\end{document}